\documentclass{aa} 

\usepackage{graphicx}

\def\deg{\hbox{$^\circ$}}   
   
\def\arcmin{\hbox{$^\prime$}}   
\def\arcsec{\hbox{$^{\prime\prime}$}}   
\def\fmag{\hbox{$.\!\!^m$}}  

\begin{document} 
 
 
\title{Shakhbazian compact galaxy groups}
\subtitle{II. Photometric and spectroscopic study of ShCG 376}

\author{Hrant M. Tovmassian
\inst{1}
\and
H. Tiersch
\inst{2}
\and
V. H. Chavushyan
\inst{3}
\and
Gaghik H. Tovmassian
\inst{4}
}

\offprints{H.M.Tovmassian}

\institute{Instituto Nacional de Astrof\'{i}sica \'Optica y
Electr\'onica, AP 51 y 216, 72000, Puebla, Pue, M\'exico
email: hrant@inaoep.mx
\and
Sternwarte K\"{o}nigsleiten, 81477, M\"{u}nchen, Leimbachstr. 1a, Germany,
email: htiersch@uni.de
\and
Instituto Nacional de Astrof\'{i}sica \'Optica y
Electr\'onica, AP 51 y 216, 72000, Puebla, Pue, M\'exico
email: vahram@inaoep.mx
\and
Observatorio Astronomico Nacional, Instituto de Astronomia,
UNAM, 22860 Ensenada, BC, M\'exico,
email: gag@astrosen.unam.mx
}

\date{Received 5July 2002 / Accepted 13 December 2002} 
 
\titlerunning{Shakhbazian compact galaxy group 376} 
\authorrunning{H. M. Tovmassian et al.}

\maketitle 
 
\abstract{
The results of the redshift measurements and of the detailed surface 
photometry in BVR of the compact group ShCG 376 are presented. The radial 
velocity dispersion, the virial mass, the total luminosity, the ${\cal M}/L$ 
ratio, and the crossing time of the group are estimated. The group consists of 
eight accordant redshift spiral galaxies. Four (or possibly five) of the group 
members have emission-line spectra. Such morphological content and the number 
of emission-line galaxies are very atypical for compact galaxy groups. There 
are signs of interaction between some members of the group. It is suggested 
that the irregular shape of the brightest galaxy No. 4 is probably due to 
interaction with other members of the group, particularly, the 
emission line galaxy No. 6 with a discordant redshift ($\Delta v = 2\,600$ km 
s$^{-1}$). It is speculated that the latter galaxy may be a infalling intruder 
to the group.

}

\maketitle

\section{Introduction}

Shakhbazian compact groups (ShCG) are richer and generally more dense groups 
of galaxies than HCGs (Hickson 1982). They were selected on the Palomar
Observatory Sky Survey (POSS) prints by {\it visual} search (Shakhbazian 1973,
Baier, \& Tiersch 1979, and references therein). In the first list of ShCGs
the groups were described as ``compact groups of {\it compact} galaxies'',
since on the POSS E prints the images of member galaxies usually lack diffuse
borders and seem to have high surface brightness. Later observations showed,
however, that most members of these groups are ordinary E or S0 galaxies.
ShCGs consist generally of 5 - 15 members; the distances between member
galaxies are typically 3 - 5 times the diameter of galaxies; the apparent
magnitudes in R of individual galaxies measured on E plates of the Palomar 
Observatory Sky Survey images are between $\approx14^m$ and $\approx19^m$;
almost all member galaxies are compact and extremely red; there are at most 1
- 2 blue galaxies in a group. In many ShCGs the space density reaches the
values of about $10^{4} - 10^{5}$ galaxies per Mpc$^{3}$. However, since these
groups are, on average, at least three times farther than HCGs (Tiersch et 
al. 1996, Tovmassian et al. 1998), and member galaxies are sufficiently faint, 
only a few of them have been investigated in detail until recently. About two 
dozen ShCGs have been observed spectroscopically (Robinson \& Wampler 
1973; Arp, Burbidge \& Jones 1973; Mirzoyan, Miller, \& Osterbrock 1975; 
Kirshner \& Malamuth 1980; Amirkhanian 1989; Kodaira et al. 1988, 1990; 
Kodaira \& Sekiguchi 1991; Lynds, Khachikian \& Amirkhanian 1990; del Olmo \& 
Moles 1991). A few years ago we commenced a spectroscopic and photometric 
study of a large number of ShCGs (Tiersch et al. 1995a, 1995b, 1996a, 1996b, 
1999a, 1999b). The number of the spectroscopically observed groups reaches at 
present about 100. The results of the study of the groups ShCG 154, 166, 328 
and 360 have been published by Tiersch et al. (2002, hereafter Paper 1). The 
reduction of observations of about 80 groups is in progress. 

During spectral and photometric study of ShCGs we encountered a unique 
group ShCG 376. In this paper we present the results of spectral and detailed 
photometric study of this group. While ShCGs consist mainly of E and S0 
galaxies, all galaxies in ShCG 376 are spirals. This group differs from other 
observed ShCGs also by its high content of emission-line galaxies.

\begin{figure*}[t]

\setlength{\unitlength}{1mm}
\resizebox{12.2cm}{!}{
 \begin{picture}(105,85)(0,0)
\put ( 10,4){}
\end{picture}} 
\caption{The image of ShCG 376 in B (left panel) and contourplots of
galaxies (right panel). North is up, east is left. Designations of
galaxies are given according to Stoll, Tiersch \& Cordis (1997).}
\label{fig1}
\end{figure*}

{\scriptsize
\begin{table*}
\caption[]{Photometric parameters of galaxies in ShCG 376.}
\medskip
\begin{tabular}{|r|ccr|ccr|ccrrr|l|}
\hline
g-xy & \multicolumn{3}{|c|}{$B$} & \multicolumn{3}{|c|}{$V$} &
\multicolumn{5}{|c|}{$R$} & type \\
\hline
  & $m$ & $b/a$ & $D$ & $m$ & $b/a$ & $D$ & $m$ & $b/a$ & $D$ &
$\alpha$ & $i$ &  \\
  &    & & $\arcsec$ &     & & $\arcsec$ &   & & $\arcsec$ & $\deg$ & $\deg$
& \\
\hline
1 &  2    &  3  &  4 &  5    & 6   & 7  &  8    &  9  & 10 & 11 & 12 & 13 \\
\hline
1 & 17.00 & 0.7 & 11 & 16.78 & 0.7 & 13 & 16.62 & 0.8 & 19 & 10 & 46 & S \\
2 & 18.85 & 0.6 & 10 & 17.83 & 0.7 & 9 & 17.45 & 0.8 & 14 & -25 & 41 & S \\
3 & 18.54 & 0.9 & 8 & 17.56 & 0.9 & 12 & 17.57 & 0.9 & 11 & 12 &  8 & S \\
4 & 15.20 & 0.9 & 50 & 15.23 & 0.9 & 23 & 15.00 & 1.0 & 37 & 32 & 19 & S/Ir \\
5 & 18.02 & 0.6 & 11 & 17.03 & 0.6 & 13 & 16.63 & 0.7 & 17 & 25 & 45 & S \\
6 & 19.47 & 0.5 & 9 & 19.10 & 0.6 &  9 & 19.12 & 0.6 & 9 & -12 & 48 & S \\
7 & 18.38 & 0.8 & 9 & 17.38 & 0.7 & 12 & 17.00 & 0.8 & 15 & 44 & 38 & S \\
8 & 17.60 & 0.8 & 11 & 16.84 & 0.8 & 11 & 16.60 & 0.8 & 14 & -63 & 40 & S \\
9 & 17.63 & 0.8 & 11 & 16.66 & 0.8 & 15 & 16.65 & 0.8 & 18 & -68 & 37 & S \\
\hline
\end{tabular}
\end{table*}
}

\section{Observations and Results}

\subsection{Direct imaging and photometry}

We obtained high-resolution images of the group ShCG 376 in BVR. The image of 
ShCG 376 in B is presented in the left panel of Fig. 1, in the right panel of 
which the isophotes of galaxies in arbitrary units are presented. They are 
drawn down to the background level. The image of the group in R is presented 
in Fig. 2. Observations were made with the 1.5 m telescope of the Observatorio 
Astron\'omico Nacional, UNAM, at San Pedro M\'artir, M\'exico. Observations in 
R were carried out in March 1998, and in B and V in May 2001. The seeing 
during observations was better than $2\arcsec$. The TEK2 CCD detector used had 
$1\,024\times1\,024$ pixels of $24$ $\mu$m squared with an image scale of 
$10.54\arcsec$/mm = $0.25\arcsec$/pxl. It covers a sky area of about 4.3 
arcmin squared. The TEK2 detector with similar parameters was used in B and V 
observations.

The reduction procedure was performed in the usual way. After subtracting the 
bias (the dark emission from the CCD itself was negligible) and the sky 
background, the image frames were divided by the evening and morning twilight 
flat fields of blank sky areas (Christian et al. 1985) to normalize the 
variations from pixel to pixel caused by different optical transmission and 
quantum efficiency. For the R colour a night sky flat field was used to avoid 
fringes. The diameters of the galaxies in ShCG 376 are quite small with 
respect to the CCD field, therefore sampling of the sky on the astronomical 
frame and its interpolation with a polynomial was possible. 

The star cluster NGC 4147 was used as a standard. It is suitable for 
photometry of galaxies because B-V colours of its stars are in the range 
$0.163 - 1.055$. The instrumental magnitudes were transformed to a 
standard BVR magnitudes using relations:

$B = b-0.31M(z)+(0.202 \pm 0.11)(B-B)+(22.040 \pm 0.038)$, \par
$V = v-0.16M(z)-(0.040 \pm 0.10)(B-V)+(23.158 \pm 0.057)$, \par
$R = r-0.11M(z)-(0.098 \pm 0.08)(B-V)+(21.914 \pm 0.048)$, \par 

where M(z) is the airmass. The magnitudes are calibrated in the Kron/Cousins 
photometric system. 

The galaxy characteristics were deduced using the SURPHOT application package 
of the MIDAS program. In general we reach the surface brightness limit 
$\mu=26\fm5$/arcsec$^{2}$ in all three bands. The magnitudes were corrected 
for galactic extinction which in the direction of ShCG 376 (R.A.=$13^h 56^m 
36^s$, Dec=$23\deg~21\arcmin~37\arcsec$) is $0\fmag094$ in B (Schlegel, 
Finkbeiner, \& Davis 1998). The measured magnitudes in B are corrected also 
for the extinction within galaxies according to $A_{B} = 0.72 \ log (1/\cos 
i)$). The corrections in V and R are calculated using the color excesses 
$E_{B-V} = 0.238 A_{B}$ and $E_{V-R} = 0.590 A_{V}$ respectively. The K 
correction is neglected because the group is relatively nearby. The estimated 
accuracy of magnitudes is about $0.06^m$. The diameters, the axial ratios, 
$b/a$, the position angles of major axis, $\alpha$, and the inclinations, $i$, 
of galaxies are measured from the $\mu=26\fmag5$/arcsec$^2$ contour.

The results of the photometry of the observed galaxies in ShCG 376 are
presented in Table 1 in the consecutive columns of which the following
information is given: 1 - the galaxy identification number; 2 - the magnitude
in $B_{26.5}$; 3 - the axial ratio, $b/a$, in B; 4 - the diameter, $D$, of the
galaxy out to the surface brightness of $26\fmag5$/arcsec$^2$ in B; 5-7 - the
latter three parameters in V; 8-10 - the same three parameters in R; 11 - the
position angle of the major axis, $\alpha$, in R at $26\fmag5$/arcsec$^2$; 12
- the inclination, $i$, in R; 13 - the galaxy type.

\begin{figure}[t] 
\setlength{\unitlength}{1mm}
\resizebox{12.2cm}{!}{
 \begin{picture}(105,85)(0,0)
\put ( 10,4){}
\end{picture}} 
\caption{The image of ShCG 376 in R filter.}
\label{fig2}
\end{figure}

We plotted curves of the isophotal surface brightness, $\mu$, versus effective 
radius, $r^{n}_{eff}$ for galaxies in ShCG 376 in colours B and R. De 
Vaucouleurs showed that ellipticals are represented by a straight line on the 
graph $\mu$ versus $r^{1/4}_{eff}$, while spirals are represented by a 
straight line on the graph $\mu$ versus $1/r_{eff}$ (Schombert 1987). Since 
ellipticals consist mainly of population II stars, we plotted the isophotal 
luminosity profiles $\mu-r^{1/4}_{eff}$ in R (Fig. 3). The spirals are 
dominated by blue population. Therefore, the isophotal surface brightness 
curves $\mu-1/r_{eff}$ were drawn in B (Fig.\ 4). The twisting profiles, 
position angle $\alpha$ versus $r^{1/4}_{eff}$ are presented in Fig. 5.

\begin{figure*} 
\setlength{\unitlength}{1mm}
\resizebox{12.2cm}{!}{
 \begin{picture}(105,45)(0,0)
\put ( 10,4){}
\end{picture}} 
\caption{Isophotal surface brightness, $\mu$, in R of galaxies in ShCG 376
versus $r_{eff}^{1/4}$.}
\label{fig3}
\end{figure*}

\begin{figure*} 
\setlength{\unitlength}{1mm}
\resizebox{12.2cm}{!}{
 \begin{picture}(105,45)(0,0)
\put ( 10,4){}
\end{picture}}
\caption{Isophotal surface brightness, $\mu$, in B of galaxies in ShCG 376
versus $1/r_{eff}$.}
\label{fig4}
\end{figure*}

\begin{figure*} 
\setlength{\unitlength}{1mm}
\resizebox{12.2cm}{!}{
 \begin{picture}(105,45)(0,0)
\put ( 10,4){}
\end{picture}}
\caption{Position angle $\alpha$, versus $r_{eff}^{1/4}$, for galaxies in
ShCG 376.}
\label{fig5}
\end{figure*}

\subsection{Spectroscopy}

Spectral observations of ShCG 376 were carried out with the 2.1 m telescope of 
the Guillermo Haro Observatory in Cananea operated by the Instituto Nacional de
Astrof\'{\i}sica \'Optica y Electr\'onica, M\'exico. The LFOSC 
spectrophotometer (Zickgraf et al. 1997) is fitted with a $600\times400$
pixel CCD with 1 arcsec/pxl. The read-out noise of the detector is 8 $e^-$. 
A set-up covering the spectral range of 4000-7100 \AA\  with a dispersion of 
5.3 \AA /pixel was adopted. The slit width was $2\arcsec$ and the effective 
instrumental spectral resolution was about 11 \AA. Nine galaxies of the group 
were observed in May 1998. The galaxy 4 and concentration 4a (Fig. 1) were 
re-observed in March 1999. In the last run we obtained spectra also for 
objects A, B, and C, images of which are somewhat elongated on the POSS 
prints, and therefore could be faint galaxies. In addition to the emission 
lines, discussed below, absorption features of H$\beta$, MgIb and NaD were 
generally seen in the spectra. For determination of redshifts the MIDAS 
package ({\it standard reduction - long} and {\it standard reduction - spec}) 
with programs therein was used. The radial velocities (RV) of galaxies are 
measured with an accuracy of about 40-60 km~s$^{-1}$. The RVs of the observed 
galaxies are presented in Table 2. They are corrected for solar motion 
($\Delta v = 300~\sin{l}^{II} \cdot \cos{b}^{II}$~km~s$^{-1}$). Objects A, B 
and C turned out to be stars.

{\scriptsize
\begin{table}[htb]
\caption[]{Radial velocities and equivalent widths of emission lines in the 
spectra of galaxies in ShCG 376.}
\label{tab1}
\medskip
\begin{tabular}{lccl|clcl}
\hline
gal.    & $v$ & EW &&  gal. & $v$ & EW \\
	& km~s$^{-1}$ 	         &&       && km~s$^{-1}$ \\
\hline

1 &	19\,680	& 10 &	5	& & 19\,890 & - \\
2 &	18\,430	& -  &	6	& & 17\,160 & 26 \\
3 &	20\,160	& 5? &	7	& & 20\,070 & - \\
4 &	20\,220	& 100 &	8	& & 19\,860 & 19 \\
4a &	20\,220	&    &	9	& & 20\,130 & 10 \\
\hline
\end{tabular}
\end{table}
}

The spectra of the observed galaxies are presented in Fig. 6. Three galaxies 
(No. 1, 4and 6) show emission line features of $H_{\alpha}$. In galaxy 8 the 
blend of $H_{\alpha}$ with $[SII](6717/31)$ is registered. In galaxy 9 the 
blend with $[NII]$ is measured. The presence of the emission line $H_{\alpha}$ 
is suspected also in the spectrum of galaxy 3. The equivalent widths of 
emission lines $H_{\alpha}$ or the blends are given in Table 1. Emission lines 
are absent in the spectra of three galaxies: 2, 5 and 7.

\begin{figure} 
\setlength{\unitlength}{1mm}
\resizebox{12.2cm}{!}{
 \begin{picture}(105,45)(0,0)
\put ( 10,4){}
\end{picture}}
\caption{The spectra of galaxies in ShCG 376.}
\label{fig6}
\end{figure}

\section{Discussion}

It is assumed that a galaxy is a member of a compact group if its RV does not
differ from the mean RV of the group by more than $\Delta v\approx 1\,000$ km
s$^{-1}$ (Hickson et al. 1992). The RVs of galaxies 2 and 6 differ from the
mean RV of other seven galaxies by $\approx1\,550$ and $\approx2\,800$ km
s$^{-1}$ respectively. Therefore, these galaxies generally should be treated as
foreground galaxies projected by chance over the group. However, in spite of 
rather high difference in RV, we suggest that the galaxy 2 is a member of the 
group ShCG 376. Consideration of Fig. 1 shows that outer isophotes of the 
central brightest galaxy 4 are extended towards galaxy 2. This extension is 
hardly a result of photographic summation of images of halos of both galaxies, 
since the basement of the extension in galaxy 4 is very wide. Such a form of 
the extension may obviously be due to a tidal interaction between both 
galaxies.

Tovmassian \& Tiersch (2001) showed that ShCGs are generally embedded in loose 
groups, members of which are distributed along the elongation determined by 
the proper members of corresponding groups. Therefore, a member of a group may 
have a RV exceeding the limit assumed for compact groups with smaller masses. 
Thus, galaxy 2 may be considered as a member of ShCG 376. The mean RV of 
eight galaxies (including galaxy 2) is $19800\pm585$ km s$^{-1}$. Then, the RV 
of galaxy 2 differs from the mean RV of the group by 1380 km s$^{-1}$.

Morphological types of galaxies in ShCG 376 are determined by inspection of
the isophotal surface profiles (Figs. 3, 4), as well as by inspection of the 
prime images of galaxies (Figs. 1, 2). None of the profiles in Fig. 3 ($\mu$ - 
$r_{eff}^{1/4}$) is straight, which is typical for ellipticals. Meanwhile all 
profiles in Fig. 4 ($\mu-1/r$) are straight up to the faint outskirts of 
galaxies. This means that all galaxies of the group are {\it spirals}. Direct 
images (Figs. 1, 2) show that galaxy 4 is an irregular or highly distorted 
spiral. The isophotes of this galaxy show a very complicated structure with a 
few irregular concentrations and clumps. The spiral arms, if any, are 
apparently distorted. This galaxy has also FIR (Tovmassian et al. 1998) and 
radio emission (Tovmassian et al. 1999). Its luminosity at 1.4 GHz is 
$1.1\times 10^{43}$ erg $s^{-1}$. The FIR luminosity in the interval 60-100 
$\mu$ is $2.6\times 10^{44}$ erg $s^{-1}$. The position of this galaxy on the 
$log ~F_{60} - log ~F_{1.4}$ graph (Fig. 4 in Tovmassian et al. 1999) shows 
that it obeys the correlation between the thermal emission from dust and the 
synchrotron radio emission from relativistic electrons found for star-burst 
galaxies.

As well as the above mentioned possible interaction of galaxy 4 with galaxy 2, 
the group demonstrates other signs of interaction between member galaxies. One 
may notice some asymmetry in the distribution of the brightness over galaxy 1: 
its bulge is somewhat shifted toward the central galaxy 4 (Figs. 1 and 2). 
Such asymmetry may be a result of interaction with the central galaxy. The 
twisting profiles (Fig. 5) show that the position angle of twisting in 
galaxies 1, 4, 7 and 8 varies with $r$. This may be considered as evidence of 
interaction between galaxies (di Tullio 1979; Kormendy 1982).    

The physical parameters of ShCG 376, deduced as in Paper 1, are presented in
Table 3. They are within the limits of corresponding parameters deduced for 
other Shakhbazian groups (Paper 1; Tiersch et al., in preparation). The masses 
of galaxies have been estimated using the V magnitudes and adopting a 
mass-to-luminosity ratio equal to 4 for spirals (Karachentsev 1987). In the 
consecutive lines of Table 3 the following information is given: line 1 - the 
redshift, $z$, weighted by masses of member galaxies; line 2 - the distance, 
$d$, of the group (H=55 km~s$^{-1}$~Mpc$^{-1}$); line 3 - the projected linear 
diameter of the group, $D$; line 4 - the radial velocity dispersion, 
$\sigma_v$, (weighted by masses of galaxies); line 5 - the virial radius, 
$R_{vir}$, of the group (weighted by masses of galaxies); line 6 - the virial 
mass; line 7 - the luminosity of the group, $L$, in solar units; line 8 - the 
mass-to-luminosity ratio in solar units, ${\cal M}_{\odot}/L_{\odot}$; and 
line 9 - the crossing time, $\tau_{c}$. 

{\scriptsize
\begin{table}[htb]
\caption[]{Physical parameters of ShCG 376.}
\label{tab3}
\medskip
\begin{tabular}{lrrcccr}
\hline
$z$ &&&& 0.0667 \\
$d$ [Mpc]       &&&& 364 \\
$D$ [kpc]	&&&& 290 \\
$\sigma_{v}$ [km~s$^{-1}$] &&&& 132 \\
$R_{vir}$ [kpc]	&&&& 111.2 \\
${\cal_M}_{vir}$ [$10^{11} {\cal M}_{\odot}$] &&&& 21.0 \\
$L$ [$10^{11} L_{\odot}$] &&&& 3.7 \\
${\cal_M}/L$ [${\cal M}_{\odot}/L_{\odot}$] &&&& 5.7 \\
$\tau_{c}$ [$10^6$ years] &&&& 348 \\
\hline
\end{tabular}
\end{table}
}

The situation is intriguing in the case of galaxy 6. Its RV is smaller than 
the mean RV of eight other galaxies by 2\,600 km~s$^{-1}$. Hence, it 
apparently should be closer to us than the group itself. However, some 
arguments allow us to suggest that this discordant redshift galaxy may be 
physically associated with the group ShCG 376. The point is that galaxy 6 is 
an emission line galaxy (ELG). The equivalent width of the H$\alpha$+[NII] 
blend in its spectrum is $\approx 25$ \AA. We estimated the probability of a 
chance projection of such a galaxy over the group ShCG 376 using the results 
of the survey of ELGs carried out by Alonso et al. (1999) who found that the 
overall density of ELGs with apparent $B$ magnitudes reaching $20^m$ (which is 
by about $0.^{m}5$ fainter than that of galaxy 6) and limiting distance 
$\approx 250$ Mpc (H=55 km s$^{-1}$ Mpc$^{-1}$) is 0.59 per deg$^2$. If galaxy 
6 is located at the distance of ShCG 376 (360 Mpc), then its absolute 
magnitude would be $M_B=-18.9$ and would be equal to the mean absolute 
magnitude of the sample considered by Alonso et al. (1999). Assuming that 
galaxies (including ELGs) are distributed in space uniformly (at least up to 
360 Mpc), we deduce by extrapolation that the surface density of ELGs until 
the distance of ShCG 376 would be $\approx2/$deg$^2$. The group ShCG 376 
occupies an area of $\approx2.5$ arcmin$^{2}$ on the sky (a triangle 
determined by galaxies 1, 8 and 9, see Fig. 1). The probability that we 
observe one ELG with $B=19.^{m}5$ projected by chance over the area of 
$\approx2.5$/arcmin$^{2}$ is equal to $P=0.0013$.

On average, only about $20\%$ of member galaxies in ShCGs are spirals
(Tiersch et al. 1996a, 1996b; Paper 1). Such a small number of spirals in ShCGs
is apparently due to the selection criteria, since groups of only {\it compact}
galaxies were sampled (Shakhbazian 1973). Therefore, ELGs among ShCG members 
are generally very rare. Contrary to the typical ShCGs, the group ShCG 376 
consists of {\it only} spirals, and the relative number of ELGs is high here. 
Hence, independent of the reasons, the group ShCG 376, because of large number 
of spirals, is unique among ShCGs. A few HCGs (e.g. HCG 16, 80, 88, 89) also 
consist of only spirals and irregular galaxies (Hickson 1994). However, each 
of the latter groups contains only four accordant redshift spirals, while ShCG 
376 has the twice number of spirals. If one takes into account that an ELG 
is ``projected'' over this unique group (one in about 100), the probability of 
a chance projection deduced above would be even smaller. Therefore, though the 
probability is estimated for a single observed case, one may suppose that 
galaxy 6 is, probably, physically associated with ShCG 376. It may be an 
in-falling intruder to the group.  

If galaxy 6 passed recently at a distance of about 10-40 kpc from the
gravitational center of the group, and its residual velocity happened to be 
orientedin the direction close to the line of sight, it could have about the 
observed value, 2\,000-3\,000 km s$^{-1}$. A disturbed forms of the central 
galaxy 4 may be a result of such passage. Moreover, as it is seen in Fig. 2, 
the outer envelope of galaxy 5 is directed exactly toward galaxy 6, while the 
orientation of the bulge is different. This may be considered as a hint that 
the accordant redshift galaxy 5 is in interaction with discordant redshift 
galaxy 6. Thus, galaxy 6 may indeed be dynamically associated with ShCG 376. 
If so, then we see the second intruder in a compact group, the first one being 
NGC 7\,318B in Stephan's Quintet (SQ) found by Burbidge \& Burbidge (1961) 
more than forty years ago. Although the suggestion of the infall of galaxy 6 
into the group is very speculative, we think that it is worth obtaining deep 
images of the group ShCG 376 and some other data to see whether there are 
direct signs of encounter of galaxy 6 with other members of the group, as in 
the case of SQ (Moles, Sulentic \& M\'arques 1997). If the collision of galaxy 
6 with ShCG 376 would be proved, then the fact mentioned by Sulentic (1987, 
1997) that the number of discordant redshift galaxies in HCGs is too large to 
be explained by a chance projection, may be due to the galaxies infalling to 
the corresponding groups with relatively high velocities, and therefore 
considered as field galaxies, not related dynamically to these groups.

\section{Conclusions}

Spectral and photometric study showed that the group ShCG 376 is very peculiar 
among about 100 ShCGs. Its all eight accordant redshift members are spirals, 
four (or possibly five) of which have emission lines in spectra. Such a
morphological content is very unusual for ShCGs even taking into account the 
selection effects. By its physical parameters (the radial velocity dispersion, 
the virial radius, the virial mass, the mass-to-luminosity ratio, and the 
crossing time) ShCG 376 does not differ from the other studied ShCGs. 

It is shown that a few galaxies in the group are probably interacting with
each other. The brightest galaxy of the group, No. 4, is the most disturbed
one.
 
We speculate that the discordant redshift galaxy 6 ($\Delta v = 2\,600$ km
s$^{-1}$) is probably colliding with the group. The apparently disturbed forms
of galaxy 4, its radio and FIR emission may be the result of the passage of
galaxy 6 through the group.

\begin{acknowledgements}
The authors are grateful to the anonymous referee for careful reading of the 
manuscript and valuable comments. HMT acknowledges the K\"onigsleiten 
Observatory for the partial financial support (accommodation) during 
June-August 2001. HT is grateful to Mr. O.Beck for private financial support.
\end{acknowledgements}

\end{document}